\documentclass[%
 reprint,
 amsmath,amssymb,
 aps,
]{revtex4-2}

\usepackage{graphicx}
\usepackage{dcolumn}
\usepackage{bm}

\begin{document}

\preprint{APS/123-QED}

\title{Characterizing Fano resonances during recollision}

\author{Graham G. Brown}
\author{Dong Hyuk Ko}
\author{Chunmei Zhang}
\email{chunmei.zhang@uottawa.ca}
\author{P. B. Corkum}
\affiliation{%
 Joint Attosecond Science Laboratory, Department of Physics, University of Ottawa, Ottawa Canada K1N 6N5\\
 National Research Council of Canada, 100 Sussex Drive, Ottawa, Canada K1A 0R6
}%

\date{\today}

\begin{abstract}
When intense light irradiates a quantum system, an ionizing electron recollides with its parent ion within the same light cycle and, during that very brief (few femtosecond) encounter, its kinetic energy sweeps from low to high energy and back \cite{01PhysRevLett.71.1994}.  Therefore, recollision offers unprecedented time resolution and it is the foundation on which attosecond science is built.  For simple systems, recolliding trajectories are shaped by the strong field acting together with the Coulomb potential and they can be readily calculated and measured \cite{02PhysRevA.49.2117, 03Torlina_2017}.  However, for more complex systems, multielectron effects are also important because they dynamically alter the recolliding wave packet trajectories \cite{04Shiner2011, 05PhysRevLett.111.233005}.  Here, we theoretically study Fano resonances \cite{06Limonov2017}, one of the most accessible multielectron effects, and we show how multielectron dynamics can be unambiguously isolated when we use in situ measurement \cite{07Dudovich2006}.  The general class of in situ measurement can provide key information needed for time-dependent ab initio electronic structure theory and will allow us to measure the ultimate time response of matter \cite{08PhysRevLett.94.033901}.  
\end{abstract}

\maketitle


Attosecond pulses generated through recollision are measured using two broad classes of methods: (i) ex situ \cite{09Paul1689} and (ii) in situ \cite{10PhysRevA.94.023825}. While we will focus on in situ techniques, we begin by briefly explaining ex situ techniques for comparison. Ex situ methods, such as photoelectron streaking, rely on the photoelectric effect to create an electron replica of the attosecond pulse by photoionizing a selected atom in the presence of an infrared pulse \cite{09Paul1689, 11Mairesse1540}. For a well-understood photoemitter, attosecond streaking fully characterizes an attosecond XUV pulse. However, from a fundamental physics perspective, underlying any streaking measurement is information about both the generating and measuring atom, including their multielectron dynamics and frequency dependent time delays due to the static structure of the residual ion (the latter is encoded in the phase of the transition moment).  These effects can be difficult to deconvolve. In contrast, in situ methods add a perturbing field to the recollision process \cite{07Dudovich2006} and observe the signal by photon emission. The relative phase between the driving and perturbing fields is used to characterize the electron continuum dynamics.  In situ methods are insensitive to the recombination dipole matrix element and associated structural effects \cite{10PhysRevA.94.023825}. That is, they are only sensitive to the recollision electron trajectory dynamics. 

The purpose of this letter is to confirm that in situ measurement provides a unique window into the effect of dynamic multielectron interaction during recollision.  We do this by investigating how multielectron Fano resonances manifest themselves within the restricted domain of attosecond in situ measurement. 

\begin{figure}
\includegraphics[width=\columnwidth]{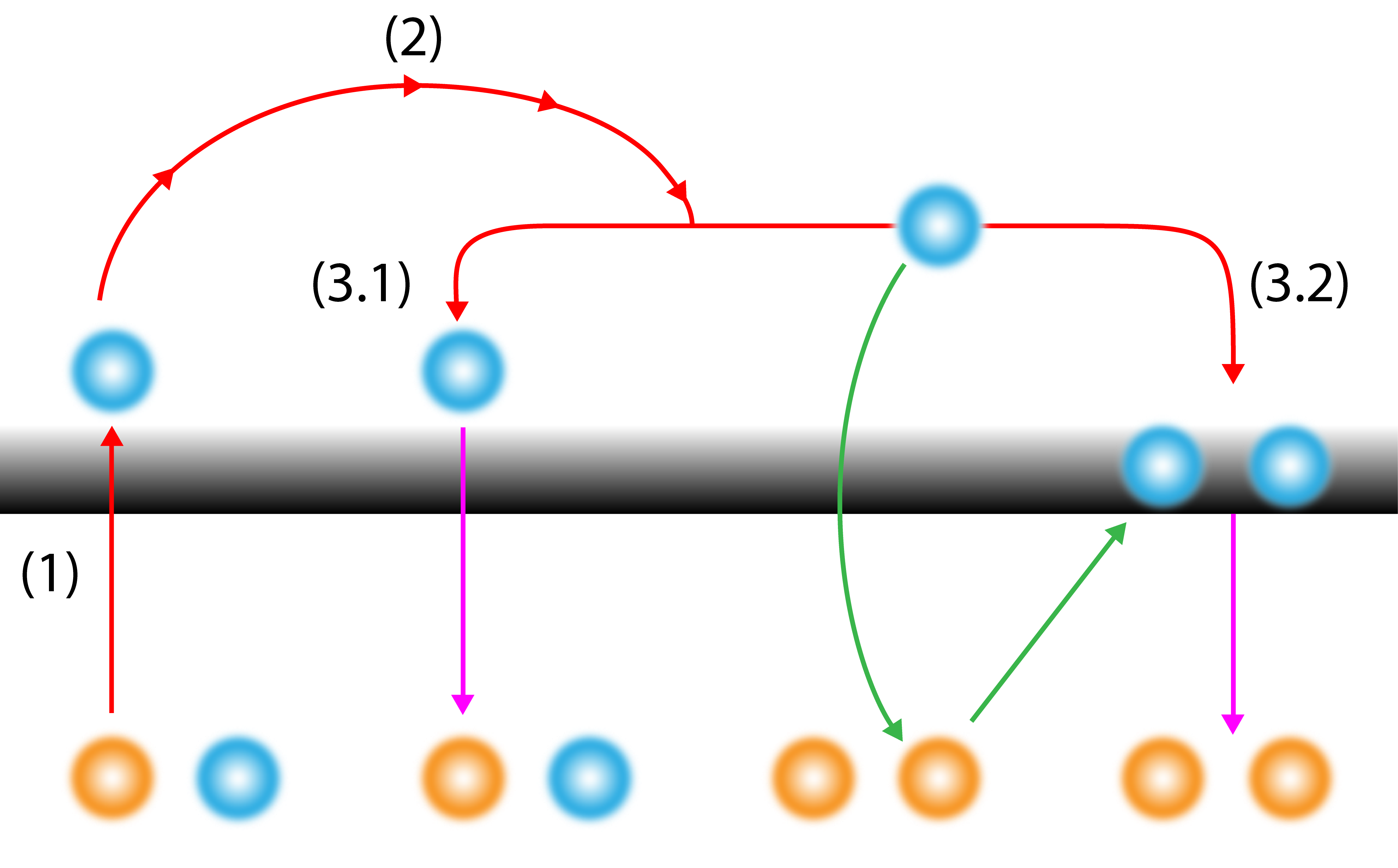}
\caption{\label{fig:cartoon}An illustration of the recollision process in a one-dimensional helium atom. (1) The electron (blue) tunnels into the continuum from the ground state, denoted by the red arrow, leaving a hole (yellow) in the ion. (2) The electron is accelerated by the field and eventually driven back towards the parent ion. The electron can recombine with the parent ion in two different ways: (3.1) the electron can recombine with the hole which was generated when it left, denoted by the left purple arrow; or (3.2) the electron can exchange energy and momentum with the remaining electron and excite it, denoted by the green arrows, subsequently recombining (right purple arrow) through a doubly-excited state embedded in the continuum (shaded grey area). The interference of these two pathways gives rise to an asymmetric resonance profile. }
\end{figure}

Fano resonances are a broad class of resonances that can occur during photoionization or photorecombination due to electron-electron interaction \cite{12PhysRevLett.94.023002, 13PhysRevA.71.060702, 14PhysRevLett.105.263003, 15PhysRevLett.108.123601}. They can result from the interference between a continuum state and a bound doubly excited state that is embedded in the continuum \cite{16PhysRev.124.1866}. These resonances represent the simplest dynamic electron-electron interaction reproducible by exact numerical methods. Within this context, there are two pathways to recombination, which are depicted in Fig. 1 and justified by the strong field model \cite{17Patchkovskii_2012} and in the Supplementary Information. The first pathway, labeled as the direct recollisional pathway (3.1 in the figure) corresponds to the well-known three-step model [1] -- the electron tunnels into the continuum (1), is propagated by the strong driving field (2), and subsequently recombines into the hole from which it left (3.1). The second pathway, labeled as correlated recollision, involves multielectron interaction. Like in direct recollision, the electron tunnels into the continuum (1); is driven by the strong laser field (2); on its return, the recolliding electron interacts with a bound electron, transferring energy and momentum and thereby exciting the ion;  and finally, the electron ion system recombines into the ground state through a doubly excited state (3.2). The interference of these two pathways gives rise to the well-known asymmetric Fano resonance profile, typically seen in photoionization spectra. 

Since autoionization is not adequately described by the approximations of contemporary ab initio theory (e.g. Hartree-Fock, time-dependent density functional theory) \cite{18PhysRevLett.66.2305, 19PhysRevA.93.063408}, we simulate a one-dimensional helium atom [20] in order to describe multielectron Fano resonances during recollision (see Supplementary Information for simulation details). The advantage of this model is that it includes quantum mechanically exact exchange and correlation effects, albeit at the expense of reduced dimensionality.  The dimensionality has no effect on the general point we wish to make. 

\begin{figure}
\includegraphics[width=\columnwidth]{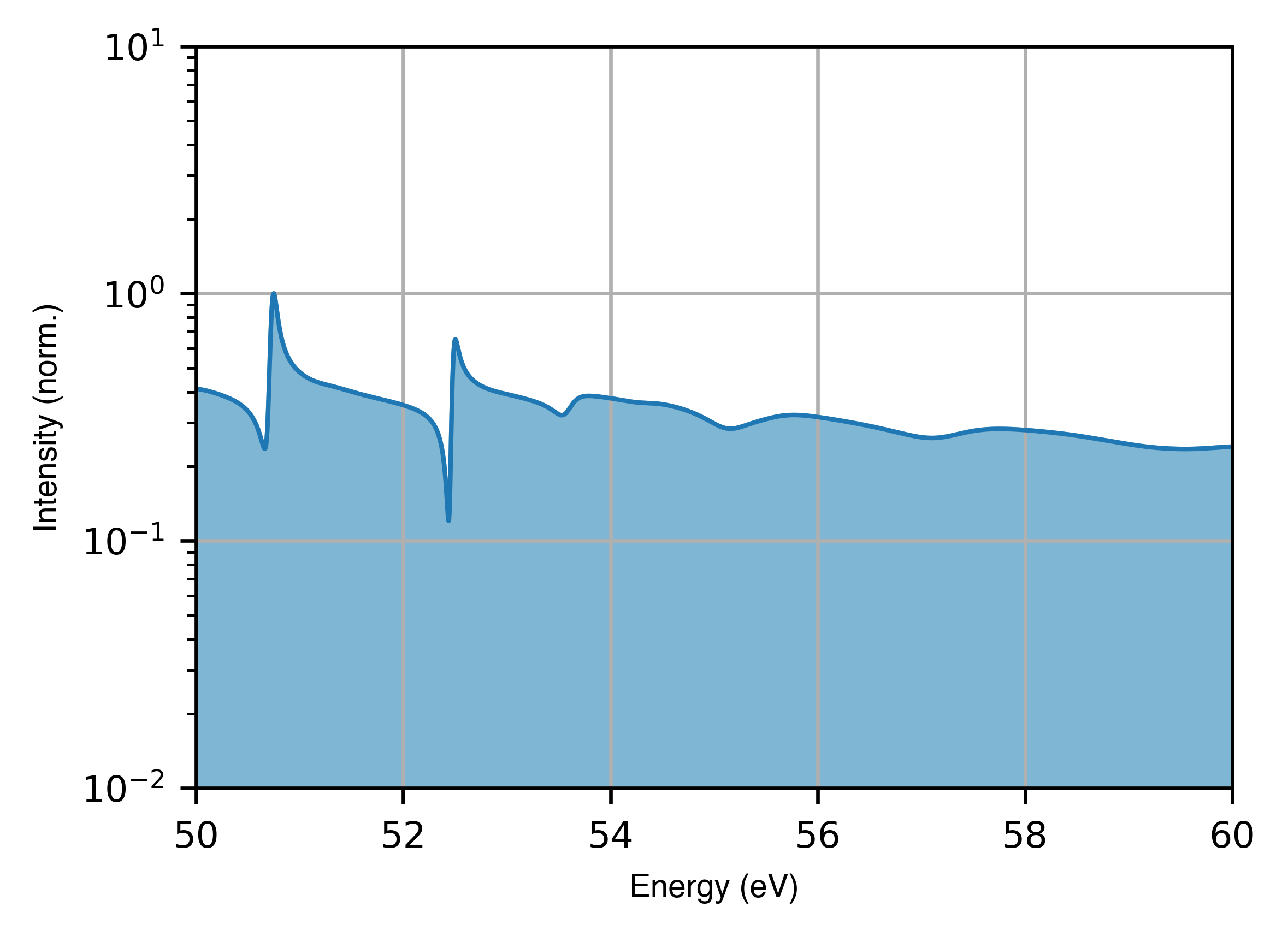}
\caption{\label{fig:cartoon}The normalized dipole acceleration spectral intensity around the Fano resonances in a simulated one-dimensional helium atom. The driving field has a wavelength of 800 nm, pulse duration of 2.5 fs and a peak intensity of  $2\times10^{14}$ W cm$^{-2}$. At energies of 50.7 eV and 52.4 eV, two sharp Fano resonant profiles dominate the recollision spectrum and, at higher energies, similar resonances are observed with softer features. }
\end{figure}

\begin{figure}
\includegraphics[width=\columnwidth]{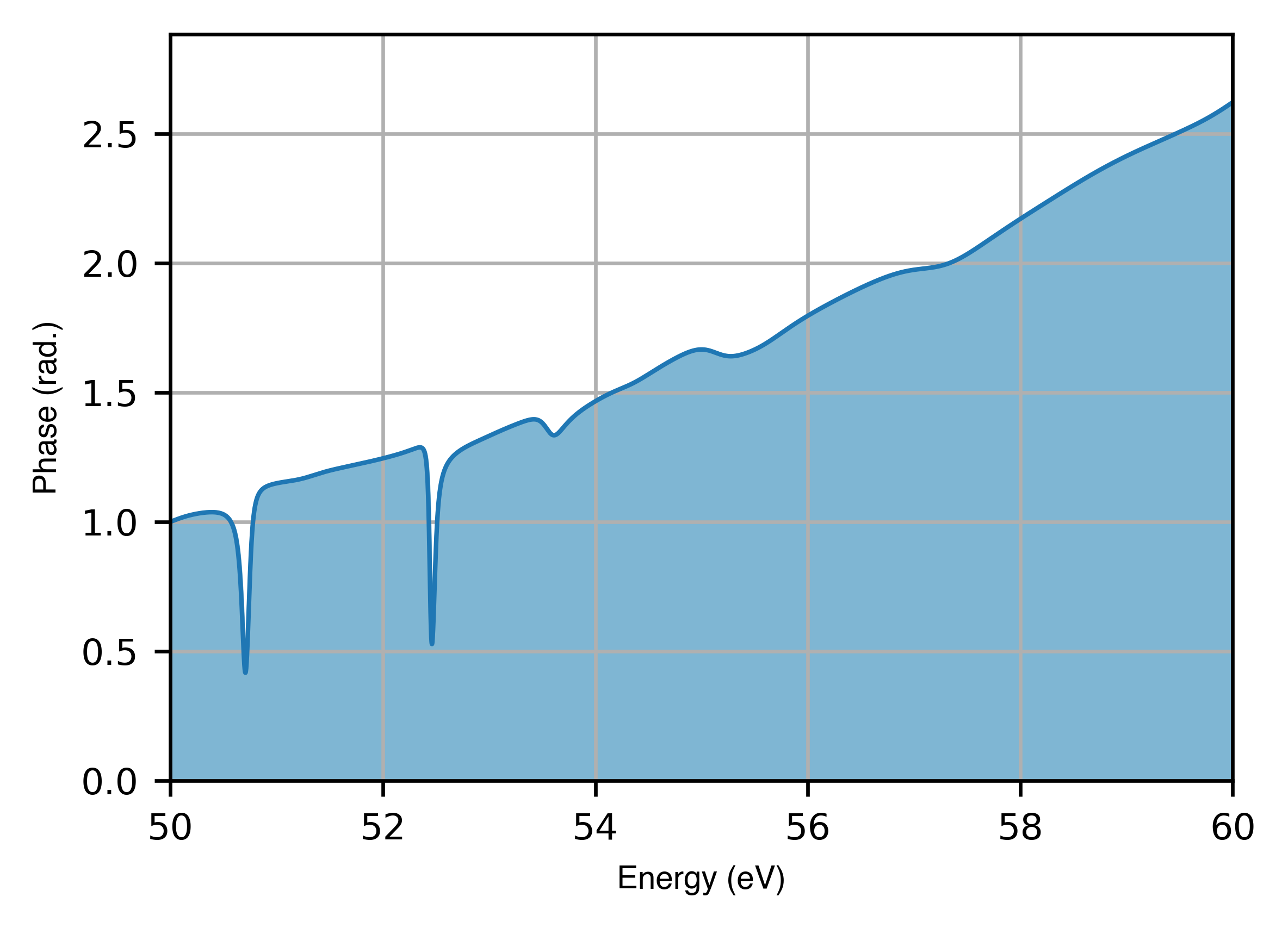}
\caption{\label{fig:cartoon} The spectral phase of the recollision spectrum around the Fano resonances in a simulated one-dimensional helium atom. The driving field has a wavelength of 800 nm, pulse duration of 2.5 fs and a peak intensity of  $2\times10^{14}$ W cm$^{-2}$. The overall shape of the spectral phase is quadratic, indicating that the direct recollision channel dominates the spectrum. Around the Fano resonances, however, the behaviour changes. At energies of 50.7 eV and 52.4 eV, the phase around the two sharp Fano resonant profiles exhibit clear deviations from the direct-recollision channel behaviour. For the higher energy resonances, the deviations are smaller in magnitude and smoother.   }
\end{figure}

The calculated dipole acceleration intensity spectrum without the perturbing field is shown in Fig. 2 for energies in the range 50 - 60 eV, where the most prominent Fano resonances in the spectrum occur. Although we are observing photorecombination, the results agree with previously published XUV photoionization results \cite{20Zhao_2012} and they are very similar in form to the photoionization spectrum measured in helium \cite{21Ott716}. The resonant structures are different for different photon energies.  The lower energy structures are sharper while the higher energy ones are softened. The amplitude and phase around a Fano resonance are highly dependent upon the matrix elements relating the ground, doubly-excited, and continuum states involved in the resonance and can vary considerably, especially in the presence of another coherent channel \cite{22Kotur2016}. 

The spectral phase of the spectrum shown in Fig. 2 is depicted in Fig. 3. The overall shape of the spectral phase is quadratic, as the spectrum is dominated by the direct-recollision channel due to the narrow linewidths of the Fano resonances. Near the resonances, however, the spectral phase deviates from the direct-recollision behaviour due to growing importance of the correlated channels. The variation of the spectral phase around the resonances at 50.7 and 52.4 eV appear more significant than the higher energy resonances and the total phase is well-described by a Fano resonance in the presence of a linearly chirped field (i.e. the direct-recollision channel). The variation of the resonance amplitude and phase profiles across the spectrum can be attributed to the variation of the recolliding electron wave packet phase. 
Within the context of photoionization, it has been shown that the shape of Fano resonances in XUV photoionization can be controlled by an infrared perturbing pulse \cite{21Ott716}. That is, a perturbing field can modulate the phase of the doubly excited state. In the closely related recollision case that we study here, the variation in the resonance profiles across the recollision spectrum is due to the variation of the continuum state phase due to the intrinsic chirp of recollision, resulting in a large variation of Fano resonance profiles in the spectrum. 

\begin{figure}
\includegraphics[width=\columnwidth]{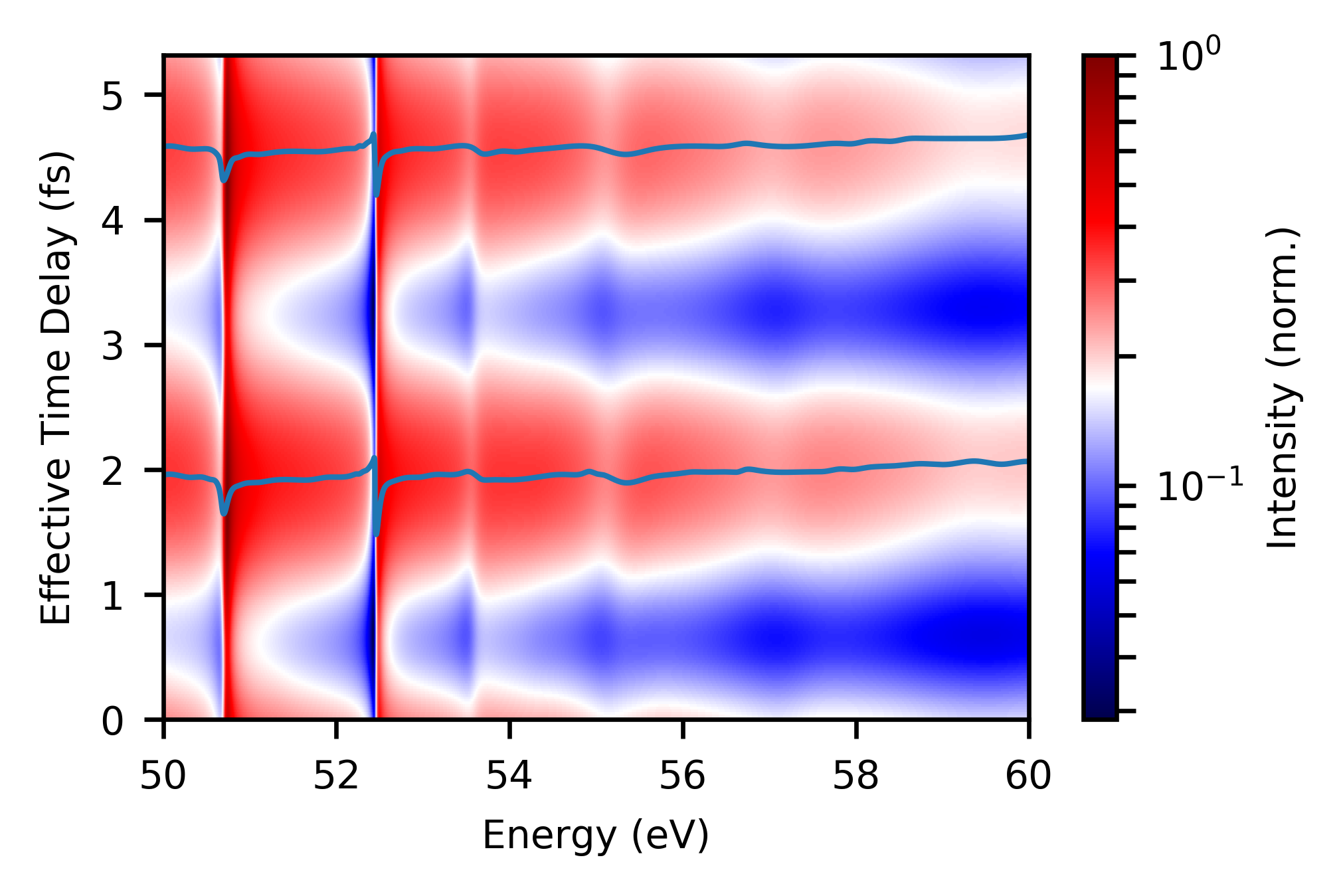}
\caption{\label{fig:cartoon}The simulated single-image in situ spectrogram for one-dimensional helium. The driving field has a wavelength of 800 nm, pulse duration of 2.5 fs and a peak intensity of  $2\times10^{14}$ W cm$^{-2}$. The perturbing field has a wavelength of 800 nm, pulse duration of 8 fs, peak intensity of $2\times10^{10}$ W cm$^{-2}$, and with a relative angle of 30 mrad with respect to the driving field. The maximizing time delay for each spectral component is plotted in blue for each grating modulation. Around each Fano resonance, the variation of the maximizing time delay deviates from the expected linear behaviour of the direct-recollision channel. These variations indicate that the electron continuum dynamics are strongly affected by the multielectron interaction leading to the observed Fano resonances. }
\end{figure}

We simulate an in situ measurement by varying the relative phase between the driving and perturbing pulse and we then superimpose the calculated intensity spectra into a spectrogram. Both pulses have frequencies of $\omega_0$ with a relative angle of 30 mrad, as explained in methods and \cite{23Ko}. The in situ measurement spectrogram for photon energies between 50-60 eV is shown in Fig. 4. The relative phase corresponding to the maximum signal for a given energy is proportional to the recollision electron group delay. For direct recollision, which exhibits a linear chirp for short trajectories, the maximizing delay varies linearly with energy. The in situ spectrogram for the one-dimensional helium atom, however, exhibits deviations from the expected linear group delay of the direct channel around each resonance. The relative delay which maximizes the recollision dipole intensity for a given energy overlays the spectrogram in Fig. 4. The lower energy resonances exhibit larger variations of group delay than the higher energy resonances, which are all quite smooth. In particular, group delay around the resonance at 50.7 eV exhibits a large negative dip, whereas the group delay for the resonance near 52.4 eV exhibits a sharp jump, indicating a comparatively positive and negative group delay for energies below and above the resonance, respectively. 

It is important to emphasize that in situ methods solely measure the electron continuum dynamics. The ground state recombination matrix element lies outside the domain of in situ methods and these methods constitute a measurement not of the emitted attosecond pulse, but of the recolliding electron wave packet group delay [10]. Typically, the linear group delay of the direct-recollision channel is interpreted as an emission time. For nonlinear group delays, however, the relationship between the spectral and temporal phase becomes quite complex \cite{24PhysRevLett.112.153001}. It is apparent from the spectrogram that, although the effect of the resonances is predominantly confined near each resonance, the entire spectrogram in the considered energy range is modulated when looking away from the maximizing time delay. This effect is stronger around the most prominent Fano resonances at 50.7 and 52.4 eV and becomes weaker at greater energies. This implies a significant reshaping of the electron wave packet. As opposed to autoionizing shape resonances that arise from the static structure of some molecular Coulomb potentials and are not observable in in situ experiments \cite{25PhysRevA.95.051401}, multielectron Fano resonances involve the dynamic reshaping of the Coulomb potential of the system’s constituent electrons due to multielectron interaction occurring during recollision. That is, they represent a dynamic change to the electronic structure of the ion and electron wave packet in response to the multielectron interaction. Due to the structural insensitivity of in situ measurement and the exact quantum mechanical treatment of the multielectron interaction simulated here, the deviations of the measurement from the expected direct-recollision channel dynamics can be ascribed solely to multielectron interaction occurring during recollision. 

In conclusion, using the simplest multielectron model that can be solved exactly using numerical methods, we have shown how multielectron effects influence high-harmonic and attosecond pulse generation. We have also confirmed that in situ measurement isolates dynamics in correlated electronic systems.   Thus, we have a new tool to study the collision-physics nature of recollision physics and an optical means of characterizing multielectron interaction occurring during recollision. This should allow collision-like measurements in transparent solids, where laser driven recollision also occurs and in situ measurements can be applied \cite{26Vampa2015}.  

We have shown that in situ measurement methods, by solely measure the continuum dynamics during electron recollision, provide an unambiguous means of studying these processes without obfuscation due to structural and recombination matrix element effects. The method we propose here is generally applicable to any process in which the momentum dynamics of the recolliding electron are altered with respect to the case of direct recollision, including autoionizing and collective resonances. This makes it a natural means to probe the ultimate time response of electronic matter \cite{08PhysRevLett.94.033901}.

Finally, due to the structural insensitivity of in situ measurement, the difference between in-situ and ex situ measurement on the same system will allow us to determine the amplitude and phase of transition moments over the full spectral region in which high-harmonic radiation can be generated.  

The authors thank Drs. Michael Spanner and David Villeneuve for many fruitful discussions.  Their demonstration that in-situ measurement methods are insensitive to photoionization time delay caused by static structure had a major impact on our thoughts.  We also  acknowledge important financial support from the US AFOSR (Grant \# FA9550-16-1-0109); the Canadian Canada Research Chairs program; the Natural Sciences and Engineering Research Council of Canada; the Canadian Foundation for Innovation; and the Ontario Research Fund.

\nocite{*}

\bibliography{characterizingFanoResonancesDuringRecollision}

\end{document}